\documentclass{ws-procs9x6}

\newcommand{\tr}{\mbox{Tr$\:$}} 
\def\lsim{\raise0.3ex\hbox{$<$\kern-0.75em\raise-1.1ex\hbox{$\sim$}}}
\def\gsim{\raise0.3ex\hbox{$>$\kern-0.75em\raise-1.1ex\hbox{$\sim$}}}

\begin{document}

\title{Free energies of static three quark systems
}

\author{ K.~H\"{u}bner, O.~Kaczmarek, F.~Karsch, O.~Vogt }

\address{Fakult\"{a}t f\"{u}r Physik\\
 Universit\"{a}t Bielefeld, D-33615 Bielefeld, Germany}

\maketitle

\abstracts{
  We study the behaviour of free energies of baryonic systems
  composed of three heavy quarks on the lattice in SU(3) pure gauge
  theory at finite temperature.
  For all temperatures above $T_c$ we find  that the connected part of the
  singlet (decuplet) 
  free energy of the three quark system
  is given by the sum of the connected parts of the free energies of $qq$-triplets (-sextets).
  Using renormalized free energies we can compare free energies in different
  colour channels as well as those of $qq$- and $qqq$-systems on an unique
  energy scale.
 }

\section{Introduction}
While existing studies on static baryonic systems focus on zero temperature 
simulations\cite{takahashi,alexandrou} or 
used maximal abelian gauge at finite temperature\cite{ichie},  
we have calculated the free
energies in different colour channels of heavy three quark
systems at finite temperature using Coulomb gauge. 
 
Here we restrict ourselves
to the analysis of equilateral triangles
above the critical temperature on a $32^3\times 8$
lattice in SU(3) pure gauge theory.
The $qq$-triplet and -sextet free energies have been calculated recently
in Ref.~4 and also by us in this work.
 \vskip -375pt
 \mbox{} \hfill BI-TP 2004/22\\
 \mbox{} \hfill August 2004\\
 \vskip 350pt

\section{Colour Channels of the Three Quark System}
The state of a three quark system as the product of the irreducible
representation of three quarks in colour space can be decomposed into
symmetry states 
\begin{equation}
  \label{eq:3q_basis}
  3\otimes 3\otimes 3  =  1\oplus 8\oplus 8^{\prime}\oplus 10,
\end{equation}
where $3$ is the irreducible representation of a quark in colour space and $1$
denotes the singlet, $8$
the first octet, $8^{\prime}$ the second octet and $10$ the decuplet state. 
The singlet is totally anti-symmetric, the first octet anti-symmetric in the
first and second, the second octet in the second and third component and the
decuplet is totally symmetric. 
 
The derivation of the representation of free energies in these colour
channels in terms of expectation values of Polyakov
loop correlation functions is similiar to
that for two quark systems\cite{nadkarni}, but more elaborate.
Denoting the Polyakov-loop at ${\bf x}_i$ by $L_i$ and $\beta = 1/T$, we find
\begin{eqnarray}
  \nonumber
  \label{eq:sing_frenergie}
  \exp\left(-\beta F_{qqq}^{1}\right)
  &=&\frac{1}{6}\big\langle 27\,\tr L_1 \tr L_2 \tr L_3 - 9\,\tr L_1
  \tr(L_2L_3) \big.\\\nonumber &&\big.\quad\, -9\,\tr L_2 \tr(L_1L_3)
  -9\,\tr L_3 \tr(L_1L_2) \big.\\
  &&\big.\quad\,+ 3\,\tr(L_1L_2L_3) + 3\,\tr(L_1L_3L_2)\big\rangle\\\nonumber
  \label{eq:ok1_frenergie}
  \exp\left(-\beta F_{qqq}^{8}\right)
  & = &\frac{1}{24}\big\langle 27\,\tr L_1 \tr L_2 \tr L_3 + 9\,\tr L_1 \tr(L_2L_3)\big.\\
  &&\quad\,\big.- 9\,\tr L_3 \tr(L_1L_2)- 3\,\tr (L_1 L_3L_2) \big\rangle\\
  \nonumber
  \label{eq:ok2_frenergie}
  \exp\big(\!-\!\beta F_{qqq}^{8^{\prime}}\big)
  & = &\frac{1}{24}\big\langle 27\,\tr L_1 \tr L_2 \tr L_3 + 9\,\tr L_3 \tr(L_1L_2)\big.\\
  &&\quad\,\big. - 9\,\tr L_1 \tr(L_2L_3)- 3\,\tr (L_1 L_2 L_3) \big\rangle\\\nonumber
  \nonumber
  \label{eq:dek_frenergie}
  \exp\left(-\beta F_{qqq}^{10}\right)
  & = &\frac{1}{60}\big\langle 27\,\tr L_1 \tr L_2\tr L_3 + 9\,\tr L_1 \tr(L_2L_3)\big.\\\nonumber
  &&\quad\,\,+9\,\tr L_2 \tr(L_1L_3) + 9\,\tr L_3 \tr(L_1L_2)\big.\\
  &&\quad\,\,\big.+ 3\,\tr(L_1L_2L_3) + 3\,\tr(L_1L_3L_2)\big\rangle.
\end{eqnarray}
With this we obtain for the average free energy of the three quark system
$F^{\mbox{\small av}}_{qqq}$ the relation
\begin{eqnarray}
  \exp\big(-\beta F^{\mbox{\small av}}_{qqq}\big)
  &=&\big\langle \tr L_1 \tr L_2 \tr L_3\big\rangle\\\nonumber
  &=&\frac{1}{27}\exp\left(-\beta F_{qqq}^{1}\right)
  +\frac{8}{27}\exp\left(-\beta F_{qqq}^{8}\right)\\
  \label{eq:average_qqq}
  &&+\frac{8}{27}\exp\big(-\beta F_{qqq}^{8^{\prime}}\big)
  +\frac{10}{27}\exp\left(-\beta F_{qqq}^{10}\right).
\end{eqnarray}
For the free energies of the $qq$-system we used the operators given in Ref.~5.
These operators as well as those defined in \eqref{eq:sing_frenergie}-\eqref{eq:dek_frenergie}
are gauge dependent and thus have to be evaluated in a fixed gauge.
We used Coulomb gauge for our calculations.

\section{\label{renorm}Perturbation Theory and Renormalisation}
Table \ref{table} summarizes the Casimirs $c_s$ found for the free
energies in the different
colour channels of the quark systems $q\bar q$, $qq$ and $qqq$. 
Using this one obtains
the leading order perturbative behaviour of the free energy in the symmetry state
$s$ as well as that of the average free energy for small distances or high
temperatures 
\begin{equation}
  \label{eq:fre_sample}
  \beta F^s(R) = c_s\frac{\alpha\beta}{R}\quad\mbox{and}\quad
  \beta F^{\mbox{\footnotesize av}}(R) = 1 + c_{\mbox{\small av}}\frac{\alpha^2\beta^2}{R^2},
\end{equation}
where $\alpha=g^2/4\pi$. 
Here $R$ is the usual Euclidian distance. 
For $qqq$-systems we restrict ourselves to equilateral triangles. In this case
$R$ denotes the 
edge length, which is an appropriate distance measure. 
In the case of the two octet free energies 
it is convenient to calculate the average of both. 
For small distances, the
average free energies behave like the left most colour channels up to
a $T$-dependent constant.
\begin{table}[ph]
  \tbl{Casimirs $c_s$ and $c_{\mbox{\tiny av}}$ for the leading order
    behaviour of $F^s$ and $F^{\mbox{\tiny av}}$.
}
{\footnotesize
  \begin{tabular}{rcccccc}
    system & average & singlet  & triplet & sextet & octet & decuplet\\
    \hline
    $q\bar q$ &$-4/9$&$-4/3$&&&$+1/6$&\\
    $qq$      &$-4/9$&&$-2/3$&$+1/3$&&\\
    $qqq$     &$-4/3$&$-2$&&&$-1/2^{\star}$&$+1$\\
    \hline
    \multicolumn{7}{l}{$^{\star}$This simple
    form only holds for equilateral triangles.}
  \end{tabular}
  \label{table}
  \vspace*{-2pt}
}
\end{table}
In the following we will show results for renormalized free energies. These are
obtained from renormalized Polyakov loops. The relevant renormalization
constants have been determined in Ref.~6 from an analysis of $q\bar q$ free
energies and can directly be used also for the $qqq$-systems.

\section{Colour Channels above $T_c$ in $qq$- and $qqq$-systems}
\begin{figure}[t]
  \centerline{\epsfxsize=2.2in\epsfbox{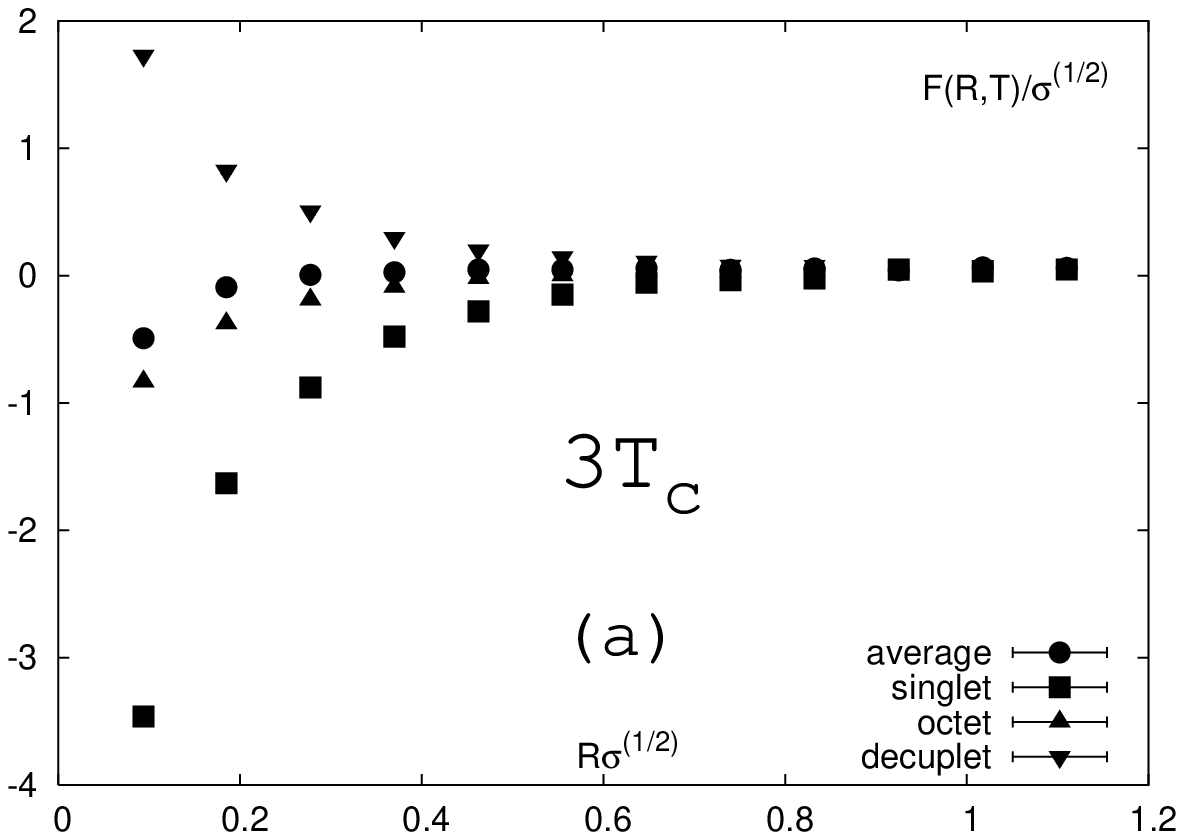}
    \epsfxsize=2.2in\epsfbox{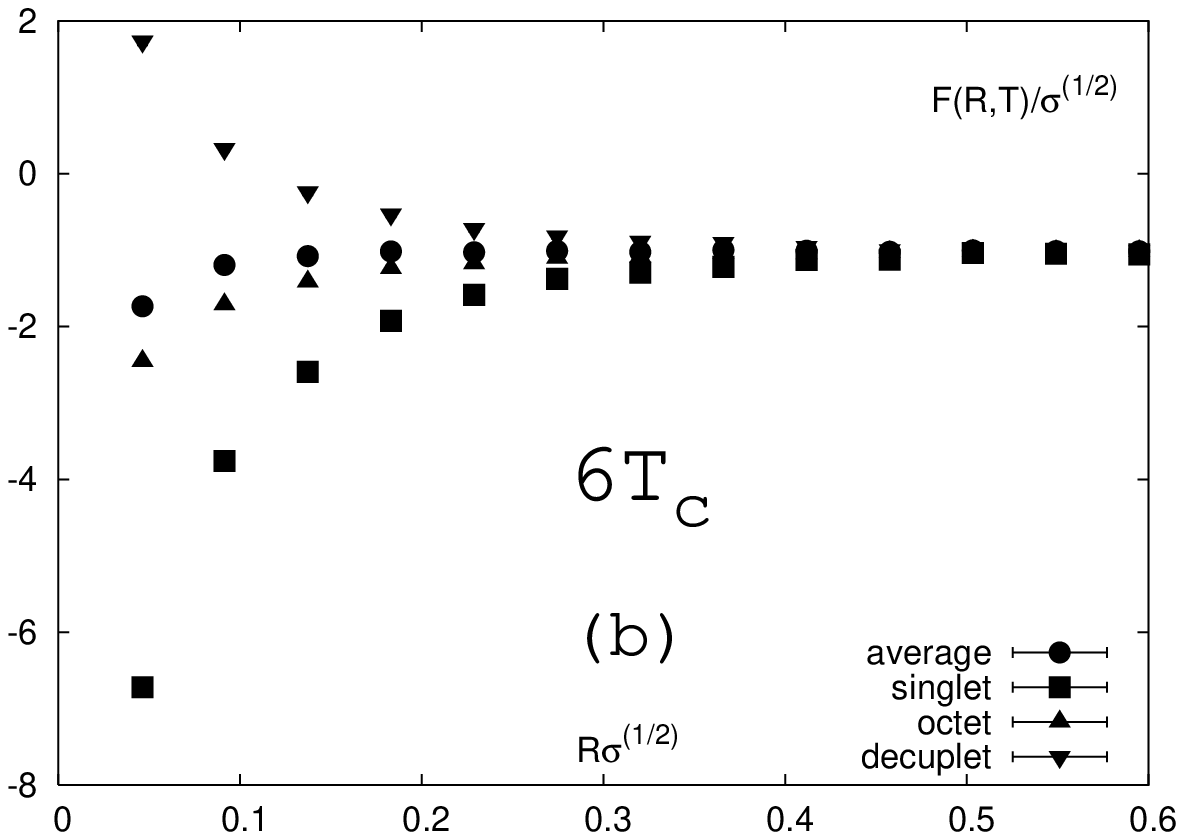}}
  \centerline{\epsfxsize=2.2in\epsfbox{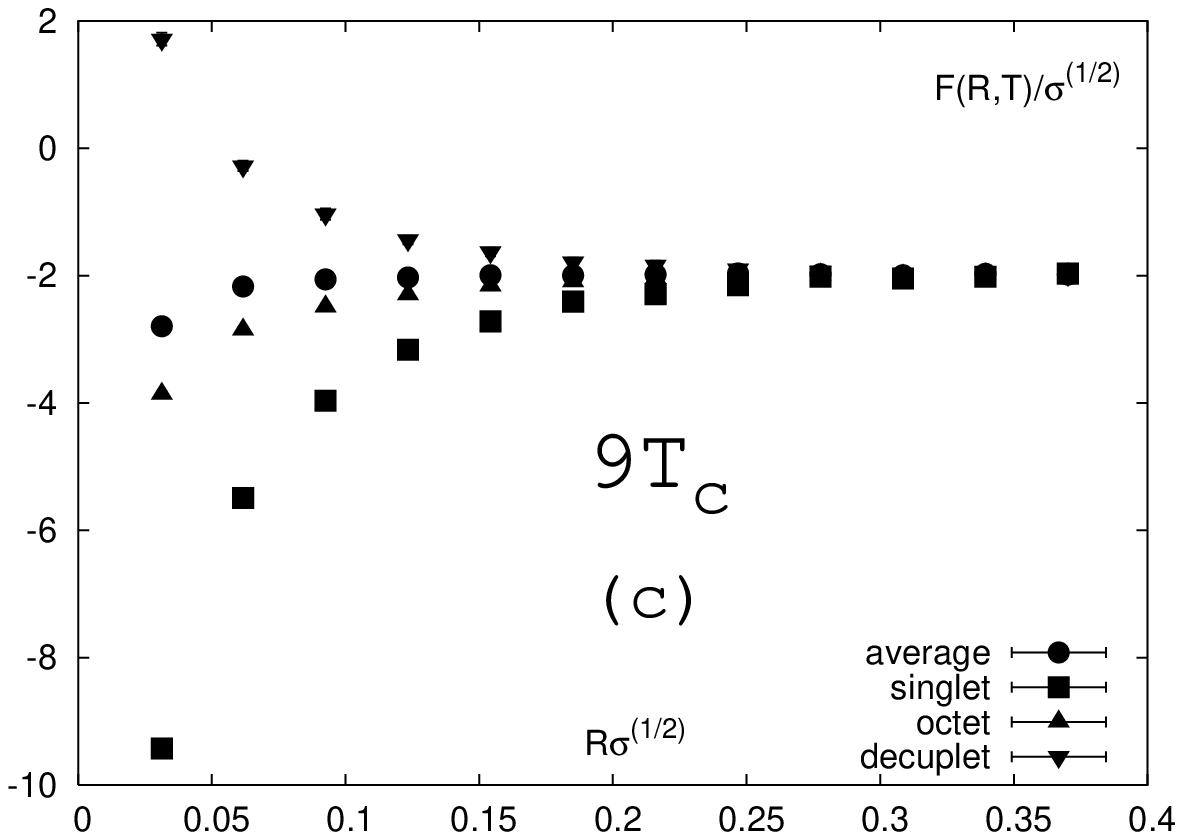}
    \epsfxsize=2.2in\epsfbox{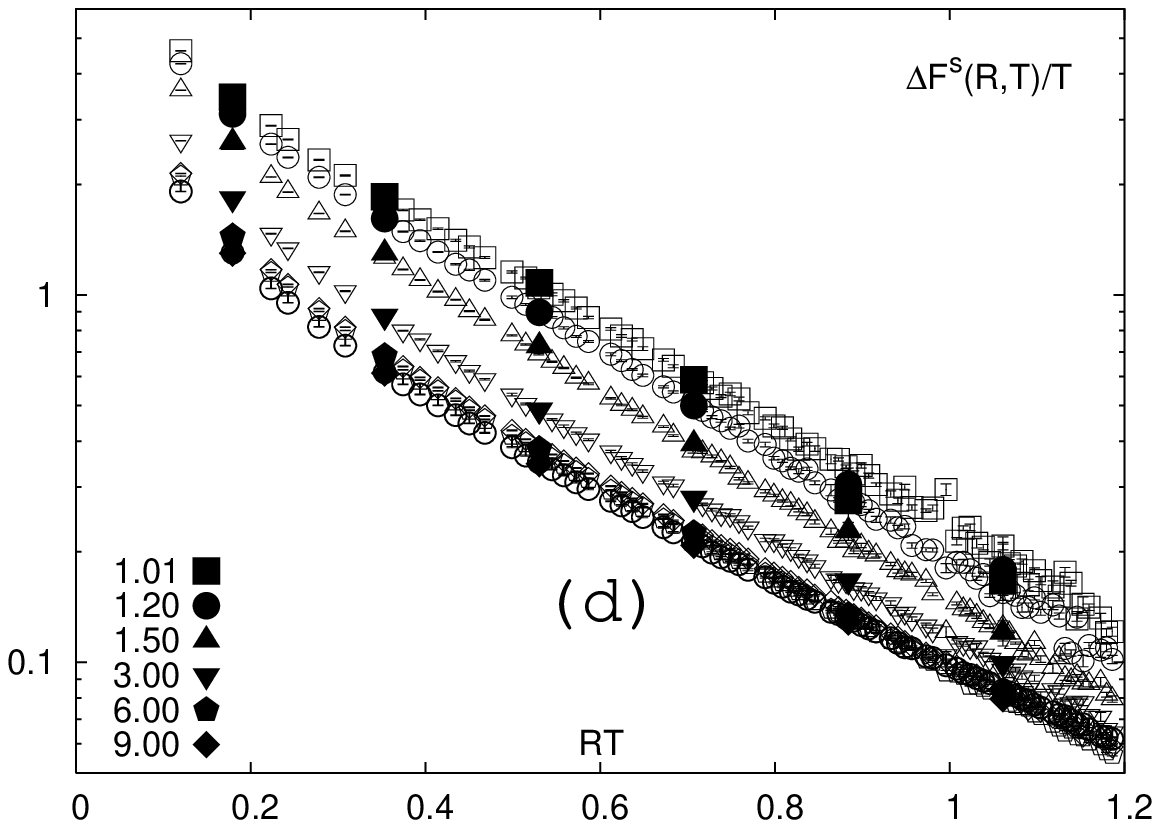}}
  \caption{(a)-(c) Free energies of the $qqq$-system in different colour
  channels for temperatures $T/T_c=3,6,9$. (d) Free Energies of the $qq$-(open symbols) and
  $qqq$-(full symbols) systems above
  $T_c$ on a $32^3\times 8$ lattice. Shown are $\Delta F_{qqq}^1(R,T)$ and
  $3\Delta F_{qq}^3(R,T)$ logarithmically.}
  \label{figure1}
\end{figure}

We compare the behaviour of the free energy in different colour
channels for temperatures $T/T_c=3,6,9$ on a $32^3\times 8$ lattice and
for equilateral triangular configurations in Figure \ref{figure1}a-c. 

One can see clearly that the singlet
free energies are strongly, the octet weaker attractive and the
decuplet free energies are repulsive. Together with \eqref{eq:average_qqq} 
this results in
weakly attractive average free energies. 
All free energies at a given temperature approach a common $T$-dependent constant
for large $R\sqrt{\sigma}$.

In Figure \ref{figure1}d we show the connected $qqq$-singlet free energies for equilateral
triangles 
\begin{equation}
  \label{eq:fre_subtracted}
  \Delta F^1_{qqq}(R,T) = F^1_{qqq}(R,T) - F^1_{qqq}(\infty,T)
\end{equation}
and compare it to the corresponding connected  $qq$-triplet free energies,
scaled by a factor of $3$. 
Over the entire $RT$-interval we find that $\Delta F_{qqq}^1(R,T)$ and
$3\Delta F_{qq}^3(R,T)$ coincide within errors for all temperatures, which
suggests that above $T_c$ the interactions of the quarks in the 
$qqq$-singlet state can be decomposed into the  pairwise interaction of three
$qq$-pairs in a triplet state.
The screening masses of both free energy channels are equal within errors.
On the other hand we find that
$F^1_{qqq}(\infty,T)=\frac{3}{2}F^3_{qq}(\infty,T)$, which shows that at large
distances the static quark sources are screened independently by a gluon cloud.
We find an analogous relation for the $qqq$-decuplet and $qq$-sextet free
energies above $T_c$, the $qqq$-octet free energies 
show, however, some small deviations.

In Figure \ref{figure2}a we show $F_{qqq}^1(R,T)$ for several temperatures
above $T_c$ obtained on a $32^3\times 8$ lattice renormalized by the
procedure mentioned in section \ref{renorm}.  
For small distances we observe that for all temperatures the $qqq$-singlet free
energies coincide, thus becoming $T$-independent.

\section{Conclusion and Outlook}
We calculated the free energies in different colour channels of heavy three quark systems at
finite temperature. 
While the connected part of the $qqq$-singlet (-decuplet) free energies are found to be
decomposable into three $qq$-triplet (-sextet) free energies for all distances
calculated above $T_c$,
the asymptotic large distance value of the free energies can be understood in
terms of three independently screened quark sources. 
The $qqq$-octet free energies show deviations.

\begin{figure}[t]
  \centerline{\epsfxsize=2.2in\epsfbox{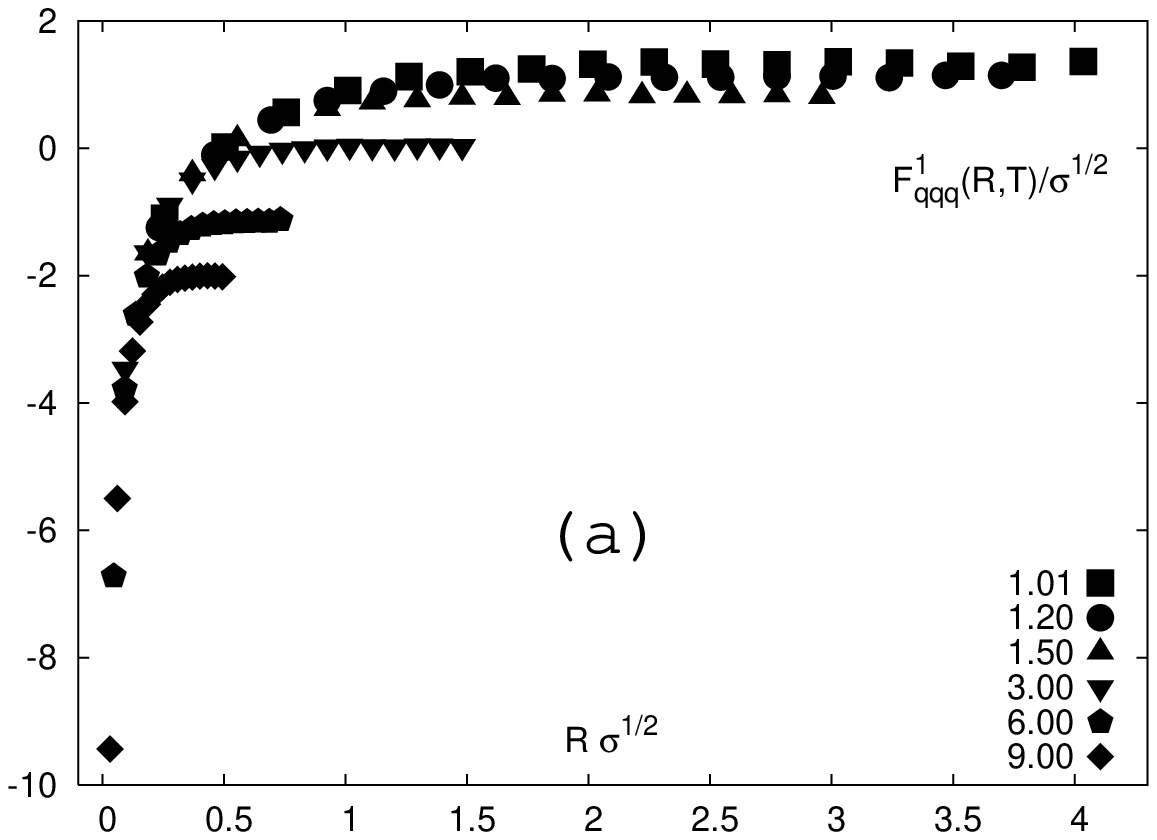}
    \epsfxsize=2.2in\epsfbox{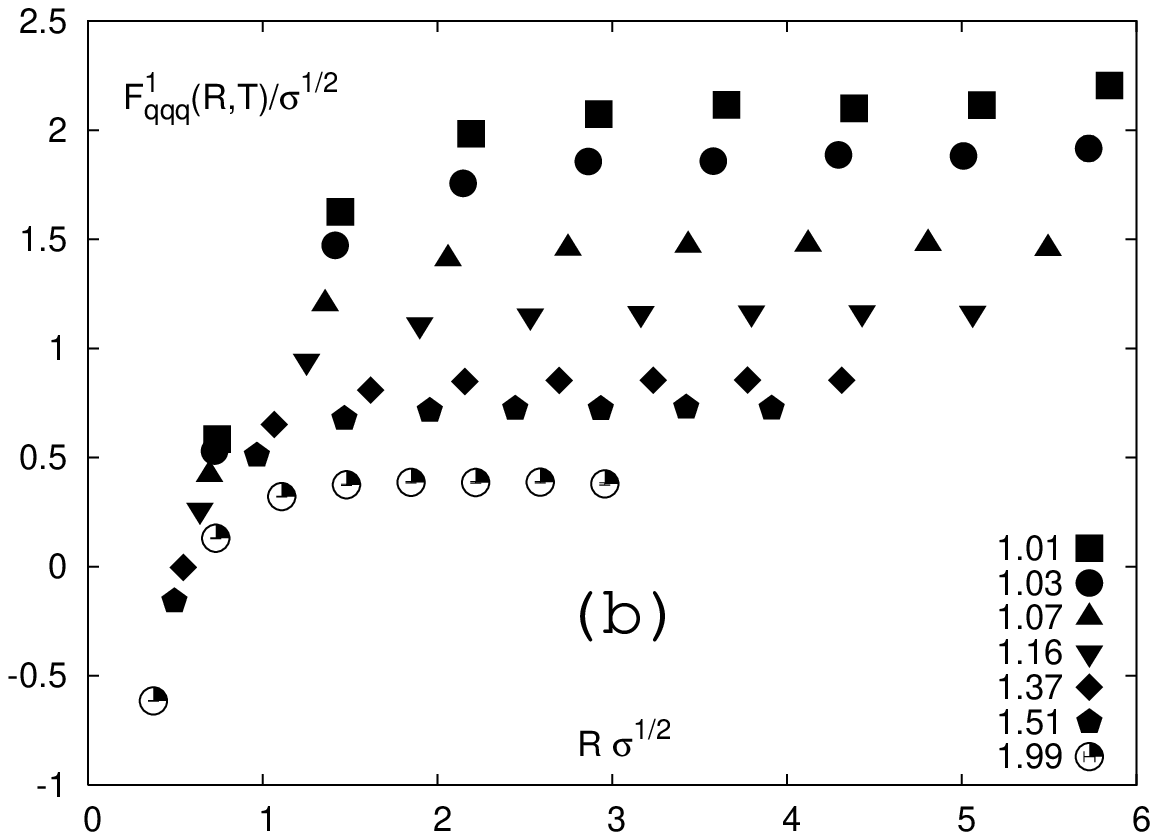}}
  \caption{
    $F_{qqq}^1(R,T)$ (a) in pure gauge on a $32^3\times 8$ lattice, (b) in 2-flavour QCD on a $16^3\times 4$
    lattice at $m/T=0.4$.   
    \label{figure2}}
\end{figure}

The approach used here for SU(3) gauge theory can also be applied to full QCD.
In Figure \ref{figure2}b we show first results for  
$F_{qqq}^1(R)$ obtained on a $16^3\times 4$
lattice in 2-flavour QCD. 
As in Figure \ref{figure2}a, the
$qqq$-singlet becomes $T$-independent for small distances.
Like the results in pure gauge simulations, we find that $qqq$-singlet (decuplet) free
energies can be decomposed into $qq$-triplet (-sextet) free energies. 

In the future we plan to perform a more detailed comparison between $qq$- and 
$q\bar q$-free energies in pure gauge and full QCD. 
Moreover, we are presently increasing the statistics for three quark free
energies below $T_c$ to be able to decide which flux tube
geometry is realised close to $T_c$. 

\section*{Acknowledgements}
This work has been supported in parts by the German Research Foundation (DFG)
under contracts GRK881 and KA1198/6-4.

\end{document}